# Distributed Dynamic Inter-Cell Interference Management for Femtocell Networks Using Over-the-Air Single-Tone Signaling

C. Yang, C. Jiang, M. Wang

*Abstract*—Femtocell networks are promising for not only improving the coverage but also increasing the capacity of current cellular networks. The interference-limited reality in femtocell networks makes interference management (IM) the key to maintaining the quality of service and fairness in femtocell networks. Over-the-air signaling is one of the most effective means for fast distributed dynamic IM. However, the design of this type of signal is challenging. In this paper, we address the challenges and propose an effective solution, referred to as single-tone signaling (STS). The proposed STS scheme possesses many highly desirable properties, such as no dedicated resource requirement (no system overhead), no near-far effect, no inter-signal interference, and immunity to synchronization error. In addition, the proposed STS signal provides a means for high quality wideband channel estimation for the use of coordinated techniques, such as coordinated beamforming. Based on the proposed STS, two distributed dynamic IM schemes, ON/OFF power control and SLNR (signal-to–leakage-plus-noise-ratio)-based transmitter beam coordination, are proposed. Simulation results show significant performance improvement as a result of the use of STS-based IM schemes.

*Index Terms*—Single-tone signaling, interference management, femtocell networks.

## I. INTRODUCTION

Femtocell networks have been gaining a significant amount of interest for improving coverage and/or increasing capacity by higher spatial reuse via cell splitting in modern wireless communication networks [1]

[2]. However, due to the unplanned and dense deployment nature and the restricted access property (a user may be denied for access to the nearest base station and forced to obtain service from a farther base station thereby the receive power from the interferer (non-serving base station) can be much stronger than that from its serving base station) of femtocell networks, interference can be so severe such that if not effectively managed the potential benefits of femtocell networks may not be fully gained [3]-[10]. Uplink interference management in two-tier femtocell networks via power control or interference avoidance is studied [3][6][8]. Spectrum allocation between femtocell and macrocell to mitigate interference is studied [4][7][9]. A beamforming codebook restriction strategy was proposed to reduce cross-tier interference [5]. Insights on interference avoidance in OFDMA femtocells are given [10]. All in all, interference management through tighter coordination among neighboring base stations to ensure quality of service (QoS) and fairness *across cells* in femtocell networks thus becomes even more crucial than that in the conventional macrocell network. One solution is to divide the resource among neighboring base stations (i.e., a base station can only use a part of the total resource). This static resource partitioning is obviously not bandwidth efficient [11]. On the other hand, resource can be better utilized by dynamic allocation on an on-demand basis.

Fast signaling between femtocell base stations is crucial for dynamic interference management (IM). However, unlike in macrocell networks where base stations and the cellular core network are inter-connected via exclusive/leased backhaul through a standard interface (e.g., the X2 interface in LTE [12]), direct connection among femtocell base stations may not be practical. A femtocell is typically connected to the cellular core network via a third party IP backhaul [13] or an Internet service provider (ISP). The delay can be significant and may vary from 10s of msec to 100s of msec. The large and yet *unpredictable* delay makes signaling through femtocell backhaul unsuitable for fast dynamic interference management in the case of bursty and delay sensitive applications. Besides the interference among femtocells, the interference of a femtocell network to the macrocell network is also a major factor that limits the wide deployment of femtocells. For example, a user who is being served by a macrocell may be severely interfered when entering the coverage of a femtocell with restricted access, resulting in degraded service. Over-the-air signaling is thus conceivably a better solution to this type of problems (where the coordination message is sent to the interfering base stations over the air by the user since the base stations that hear the message are the strongest interferers to this user.) by providing a short-term, fast interference relief.

However, over-the-air signaling poses its own unique challenges. First, over-the-air signaling does not come for free. It requires considerable precious wireless resource for the reliable transmission of the coordination message; second, it requires deep coverage since the target receivers for the coordination message are the *neighboring* base stations instead of the *serving* base stations unlike most existing signaling schemes that are designed only for communicating to the nearest serving base stations. This requires the signal to have sufficiently large processing gain and power which also means more resource; Third, due to the broadcast nature of over-the-air signaling, conventional *random* access signaling [14]-[16] suffers from interference among signals sent by different users, which is commonly referred to as the near-far effect [17]-[19]. Since the target receivers for the current type of application are the neighboring base stations, the near-far effect is even more damaging. In today's cellular systems (e.g., WCDMA and LTE), CDMA (Code Division Multiple Access) signaling and its variants are used for random access, such as the random access signals used in uplink random access channels (RACHs) either in a CDMA system (e.g., WCDMA) or an OFDMA (Orthogonal Frequency Division Multiple Access) system (e.g., LTE) [20]-[22]. This signal (RACH) is used by users to request channel resources from its serving base station. To minimize the near-far effect, the transmit power must be carefully managed. The user first determines the minimum transmit power according to the received signal strength of the downlink pilot signal and gradually increases its transmit power at each failed attempt in order to avoid blanking out other access signals from different users within the same cell. This scheme clearly does not apply to the current application since the recipients of the coordination message are the neighboring base stations. Maximum power is typically required to transmit the coordination message to ensure sufficient coverage which inevitably blocks its serving base station from hearing the coordination signals from the users of other cells.

The main contributions of this paper are (1) to propose and analyzed a single tone over-the-air signaling scheme for dynamic inter-cell interference management (IM) in a femtocell network; and (2), to propose two IM schemes based on the proposed signaling scheme: an ON/OFF power control scheme and a signal-to-leakage-plus-noise-ratio (SLNR)-based transmitter beam coordination scheme. The difference between the work in this paper and the related work in the existing literatures is that previous works mainly focus on the theoretical analysis, centralized schemes and insights on interference management while this paper proposes practical, distributed schemes for IM in femtocell and gives some theoretical results. We will focus on the <u>*downlink*</u> IM scenario in which *the interference coordination request message is broadcast on the <u>uplink</u> by the user to the interfering base stations*. Since the base stations that hear the message are the major interferers, only these base stations need to respond to the message by participating in the coordination.

The remainder of this paper is organized as follows. Section II gives a detailed description and an analysis of the proposed single-tone signaling scheme. A concrete example of the STS design and the performance of STS are given. Section III describes two STS-based IM schemes, ON/OFF power control and the SLNR-based transmitter beam coordination. Section IV concludes this paper.

*Notations*: $(\cdot)^{-1}, (\cdot)^{T}, (\cdot)^{H}$ and $\mathbb{E}(\cdot)$ denote inverse, transpose, conjugate transpose, and expectation, respectively. $\|\cdot\|^2$ represents the Frobenius norm. $\mathrm{I}_N$ is the $N \times N$ identity matrix.

## II. SINGLE-TONE SIGNALING

### A. *Single-Tone Signaling*

In the proposed single-tone signaling (STS) design, a large fraction, if not all, of the energy in an OFDM (Orthogonal Frequency Division Multiplexing) symbol is transmitted on a *single* OFDM subcarrier. No energy is transmitted on any other subcarriers of the current OFDM symbol and no information is modulated onto the energized subcarrier (i.e. amplitude and/or phase). The energized subcarrier is referred to as the STS signal *tone*. It is the index of the energized subcarrier that contains information. That is, which subcarrier of this OFDM symbol is energized depends on the content of the message that the STS signal carries.

The coordination message carried by STS signal for IM is denoted as *m*, which can be further represented by $K$ information symbols, $\mathbf{u} = (u_1, u_2, \ldots, u_K)$, where $0 \leq u_k \leq D-1$ for $1 \leq k \leq K$, or more precisely,

$$m(D) = u_K D^{K-1} + u_{K-1} D^{K-2} + \ldots + u_2 D + u_1. \tag{1}$$

where $u_k$, $1 \leq k \leq K$, is the *index* of the energized subcarrier (i.e. the index of the STS signal tone), $D$ $(\leq S)$ is the total number of subcarriers of an OFDM symbol used for STS transmission, and $S$ is the total number of subcarriers in an OFDM symbol. Without loss of generality, we assume $D = S$ in this paper. Hence $K$ OFDM symbols are needed to transmit the message *m*, as shown in Fig. 1.

Note that an STS signal is not to be confused with a frequency hopped signal. In frequency hopping, the tone positions/indices are *predetermined* by a sequence known to both the transmitter and the receiver; the tone position itself therefore does not contain any information and the information is modulated onto the amplitude and/or phase of the tone via, e.g., QPSK (Quadrature Phase Shift Keying) or QAM (Quadrature Amplitude Modulation). While for an STS signal, the tone is not modulated with information. Instead the information is embedded in the positions/subcarrier indices of the tones.

The choice of this type of signal has the following advantages: First of all, unlike the most commonly used CDMA signals for random access [14]-[16], the STS signal does not suffer from the near-far effect among STS transmissions from different users. This is because: 1) If some of the STS signal tones from different users do happen to be on the same subcarrier, they simply add together like multi-paths (since the STS tones are not modulated therefore they share the same waveform and are thus not tone-distinctive) and are absorbed by cyclic prefix [23]; 2) If the STS signals from different users are transmitted on different subcarriers, they don't present interference to each other since all the OFDM subcarriers are orthogonal as long as the frequency between the transceivers is perfectly synchronized. Indeed, if the frequency offset is present in practice between the transceivers, the OFDM subcarriers are no longer orthogonal. However, since the tones are not distinctive between STS signals, the tone energy leakage from the neighbor subcarriers of different users to the current subcarrier, as a result of non-orthogonality, adds together with the tone of the current subcarrier just like multi-paths similar to the case in 1). This is not the case though for regular OFDM signals where the information modulated on the subcarriers from the weak users can be destroyed by the leakage from the neighboring subcarriers from the strong users, causing near-far effect when frequency offset is non-zero. This means that regular OFDM signals under imperfect synchronization also suffer from the near-far effect. This is why uplink power control is necessary even for OFDMA systems [21][24]. This unique property of STS means that the STS signal tones from different users do not interfere with each other. Strong STS signals (from local cell users) will thus not block weak STS signals (from neighboring cell users). This property is crucial for this particular application since the target receivers of the signal are the neighboring base stations rather than the serving base station.

Second, an STS signal is a narrow-band constant-profile sinusoidal waveform, therefore, has lower peak-to-average power ratio (PAPR) than regular OFDM signals [25]-[28], and allows for a higher power amplifier setting and thereby deeper coverage.

Third, since the transmit energy is concentrated on one single subcarrier of an OFDM symbol, the STS

signal tone is much stronger than a regular data tone therefore is easy to be detected by the receivers (the interfering base stations) even under strong interference environment. As a result, STS transmission can be overlaid with other users' uplink data transmission. If such a strong tone that reaches a base station becomes hard to be detected, this base station will not likely cause significant interference to the sender, i.e., the sender (the user) must be out of the interference range of the base station. On the other hand, the interference to the uplink data traffic is also concentrated on one subcarrier of an OFDM symbol. This *isolated* interference can be most effectively removed by the data decoder, thereby causing minor impact to data decoding since a decoder is most effective in removing *isolated* erasures/errors. This means that an STS signal can be transmitted without designated system resource. Therefore, other users don't need to clear the subcarriers (not transmitting on the subcarriers) for other users' STS signals in advance. Hence no overhead is incurred. This is in contrast to the conventional random access signaling scheme where a continuous chunk of the time and frequence resource has to be set aside for random access signaling using, for example, PN (Pseudorandom Noise) or ZC (Zadoff-Chu) sequences based on CDMA technology [17]-[19],[29].

Last, the STS signal provides strong wideband pilots/reference signals for accurate channel estimation in a TDD (Time Division Duplex) system allowing interference mitigation techniques to be applied for further interference minimization. This is because the STS tones are not information-modulated, they can be served as pilots/reference signals for channel estimation. The STS signal tones are strong (carries the whole OFDM symbol energy), therefore the high quality of the channel estimate. The STS signal tones are spread across the whole band, therefore, the wideband property of the channel estimate. We will see the application in Section III.

Despite the advantages of STS signal, however, STS tones can also be prone to errors due to fading (falsely detected tones or missed detection of tones). Furthermore, even under perfect detection, i.e., no falsely detected tones or missed tones, simultaneous transmissions of STS signals from multiple users may confuse the receiver due to the indistinctive nature of the STS tones. For example, if more than one STS tones from more than one users are detected in one OFDM symbol, the receiver (i.e., the base station) will not be able to distinguish the STS tones from different STS signals sent by different users due to the fact that STS tones are not information-modulated as earlier stated, causing ambiguity at the receiver. For the example in Fig. 2, the receiver may interpret one of the signals as $[u_1 \; v_2 \; v_3 \; u_4 \; ... \; u_K]^T$, where $[u_1 \; u_2 \; ... \; u_K]^T$ and $[v_1 \; v_2 \; ... \; v_K]^T$ are actually the original STS signals from user 1 and user 2, respectively. The number of such possible combinations is $G^K$ where $G$ is the number of simultaneous transmitted STS signals ($G = 2$ in Fig. 2). That is, although the tone-indistinctive property of the STS prevents inter-STS signal *tone* interference as discussed in the previous statement, it results in ambiguity at the receiver.

In addition, since the target receivers of the STS signal are multiple different neighboring femtocell base stations that may not be perfectly synchronized, the frequency offset between the transmitter and receiver may cause the received STS tones shift to different subcarriers resulting in a different STS signal.

Hence it is necessary to provide the STS signal with an ambiguity resolution capability and a certain degree of error protection. We therefore employ a special transform matrix in Galois field to transform the original non-binary information symbols $\mathbf{u} = [u_1 \; u_2 \; ... \; u_K]^T$ into a code word $\mathbf{c} = [c_1 \; c_2 \; ... \; c_K]^T$ with code rate $(N, K)$, where $N \geq K$ and $0 \leq c_n \leq D - 1$ for $1 \leq n \leq N$. Let $\alpha$ be a primitive number in Galois field, $GF(D)$, the $K$ information symbols $\mathbf{u} = [u_1 \; u_2 \; ... \; u_K]^T$ are encoded into $N$ coded symbols via the following transformation:

$$\mathbf{c}^T = \mathbf{Z} \begin{bmatrix} 0 & \mathbf{u}^T & 0 & \cdots & 0 \end{bmatrix}^T \qquad (2)$$

where

$$\mathbf{Z} = \begin{bmatrix} 1 & 1 & \cdots & 1 \\ 1 & \alpha^{\frac{D-1}{N}} & \cdots & \alpha^{\frac{D-1}{N}(N-1)} \\ \vdots & \vdots & \ddots & \vdots \\ 1 & \alpha^{\frac{D-1}{N}(N-1)} & \cdots & \alpha^{\frac{D-1}{N}(N-1)(N-1)} \end{bmatrix} \quad (3)$$

The zero element before $\mathbf{u}^T$ in (2) is inserted on purpose for frequency error protection as will be explained later. Clearly, $0 \leq c_n \leq D-1$ for $1 \leq n \leq N$. It can be shown that the resulting coded STS signal $\mathbf{c} = [c_1 \ c_2 \ ... \ c_N]^T$ achieves the largest possible code minimum distance for any linear code with the same encoder input and output block lengths (or *maximum distance separable*).

*Proposition 1: The coded STS signal is maximum distance separable.*

*Proof*: From (2) each code symbol of code word $\mathbf{c}$ can be written as

$$\begin{aligned} c_n &= \sum_{k=1}^{K} u_k \alpha^{\frac{D-1}{N}(n-1)k} \\ &= \alpha^{\frac{D-1}{N}(n-1)} \sum_{k=1}^{K} u_k \left( \alpha^{\frac{D-1}{N}(n-1)} \right)^{k-1}, \quad 1 \leq n \leq N \end{aligned} \quad (4)$$

Since a polynomial of order at most $K-1$ can have at most $K-1$ zeros, the number of non-zero code symbols in $\mathbf{c} = [c_1 \ c_2 \ ... \ c_N]^T$ is thus no less than $N-K+1$. This means that the minimum distance of $\mathbf{c}$ is at least

$$d_{\min} \geq N - K + 1 \quad (5)$$

However, by the Singleton bound, the minimum distance for any linear code also satisfies

$$d_{\min} \leq N - K + 1 \quad (6)$$

It can be readily observed from (5) and (6) that the minimum distance of $\mathbf{c}$ must be

$$d_{\min} = N - K + 1 \quad (7)$$

That is, $\mathbf{c}$ achieves the largest possible code minimum distance for any linear code. We then conclude that the STS signal is maximum distance separable. □

Like any other maximum distance separable code, the coded STS signal can correct any combination of false tone detections and missed tone detections as long as [30]

$$2\varepsilon + v \leq N - K \tag{8}$$

where $\varepsilon$ is the number of false tone detections and $v$ is the number of missed tone detections. The benefit of the maximum distance separable property of the STS signal will be seen in the following discussions.

The value of the code symbol ($c_n$, $1 \leq n \leq N$) corresponds to the *index* of the sub-carrier on which energy is transmitted. That is, only one tone is energized per OFDM symbol, and the *index* of the tone is dependent on the value of the code symbol. Either all or partial of the total (maximum) energy in an OFDM symbol is transmitted on a single subcarrier depending on the desired coverage range of the coordination scheme. Fig. 3 shows an example of the transmission of the coded STS signal. We will show in the following that the coded STS signals from different users can be *disambiguated* at the receiver (i.e., the interfering base station) such that STS tone sequences, such as $[c_1 \ d_2 \ d_3 \ c_4 \ ... \ c_N]$ or any combination of such, can be identified as invalid sequences (which does not correspond to a valid STS signal), where $[c_1 \ c_2 \ ... \ c_N]$ and $[d_1 \ d_2 \ ... \ d_N]$ are the original coded STS signals from user 1 and user 2, respectively.

## B. STS Signal Processing at the Receiver

STS signal tone detection is the first step of STS signal processing at the receiver and can be done by simply looking for a subcarrier with significantly higher energy than its neighbors. After the detection of STS tones, the receiver obtains a set of subcarrier indices on which STS tones are detected on every OFDM symbol with certain false detection tones as well as missed detection tones. By applying, for example, maximum likelihood decoding, the receiver is able to recover the original symbols.

In the presence of multiple STS signals, the STS signals from different users may overlay on top of each other causing potential ambiguity at a receiver (cf., Fig. 3). Indeed, $G$ STS signals with code rate $(N, K)$ can coexist without causing decoding ambiguity as long as the inequality, $K \leq \left\lceil \frac{N}{G} \right\rceil$, is satisfied. This important property can be formally stated by the following proposition:

*Proposition 2:* Assume $G$ $(G \leq D^K)$ distinctive STS signals coded on GF(D) with code rate $(N, K)$, $N \geq K$, are simultaneously received. Under perfect tone detection, all G coded STS signals can be decoded to the original information symbols without ambiguity, if

$$K \leq \left\lceil \frac{N}{G} \right\rceil \tag{9}$$

*is satisfied.*

*Proof:* Consider $G$ $(G \leq D^K)$ distinctive STS signals coded with rate $(N, K)$ on $GF(D)$ are simultaneously received on $N$ OFDM symbols, free of tone detection error. Now arbitrarily select $N$ number of the detected tones, each from one of the $N$ different OFDM symbols. We maintain that

1) There are at least $\left\lceil \frac{N}{G} \right\rceil$ STS tones out of the $N$ selected tones coming from the *same* STS signal among the total number of $G$ STS signals. This is the direct outcome of the pigeonhole principle [31].

2) For an STS signal with a code rate of $(N, K)$, a minimum number of $K$ STS tones are sufficient to distinguish one STS signal from another. This is sustained by the fact from proposition 1 that the coded STS signals are maximum distance separable. Indeed, since the minimum distance between STS signals is $d_{\min} = N - K + 1$ (cf., (7)), i.e., the number of different tones between any two STS signals is at least $d_{\min}$, or the number of the same tones between any two STS signals is at most $N - d_{\min} = K - 1$. Therefore a minimum number of $K$ STS tones are sufficient to determine an STS signal.

From 1) and 2), it is clear that among the $N$ STS tones, each selected from the $N$ individual received OFDM symbols, there are at least $\left\lceil \frac{N}{G} \right\rceil$ tones belonging to one of the $G$ STS signals. With these

$\left\lceil \dfrac{N}{G} \right\rceil \geq K$ tones, we can uniquely determine the corresponding STS signal. We therefore conclude that if $\left\lceil \dfrac{N}{G} \right\rceil \geq K$, i.e., (9), is satisfied, the $G$ STS signals can be uniquely separated from each other without ambiguity. That is, the receiver will not falsely detect any STS signals other than the $G$ STS signals. □

For the example in Fig. 4, two $(G = 2)$ STS signals of code rate (11, 2) on $GF(32)$,

$$\mathbf{c} = [18\ 25\ 7\ 2\ 30\ 21\ 27\ 25\ 23\ 5\ 12]^T \tag{10}$$

and

$$\mathbf{d} = [9\ 25\ 18\ 11\ 13\ 27\ 18\ 31\ 16\ 31\ 4]^T \tag{11}$$

are transmitted, carrying messages $\mathbf{u} = [5\ 1]^T$ and $\mathbf{v} = [10\ 5]^T$, from user 1 and user 2, respectively, and received by a receiver (base station). Since the STS tones are not distinctive, a receiver is not able to tell which tones belong to which STS signal. For example, in Fig. 4, the receiver cannot tell if tone 18 belongs to the STS signal $\mathbf{c}$ in (10) or to $\mathbf{d}$ in (11) and likewise for tone 9 on OFDM symbol 1. Similarly, the receiver cannot tell if tone 21 belongs to $\mathbf{c}$ or to $\mathbf{d}$ (and likewise for tone 27) on OFDM symbol 6. It thus seems that the receiver can arbitrarily interpret the received signals as

$$\mathbf{c}' = [9\ 25\ 7\ 2\ 30\ 27\ 27\ 25\ 23\ 5\ 12]^T \tag{12}$$

and

$$\mathbf{d}' = [18\ 25\ 18\ 11\ 13\ 21\ 18\ 31\ 16\ 31\ 4]^T \tag{13}$$

or any such combinations. However, out of the total $2^{11} = 2048$ such combinations, Proposition 2 guarantees that only two are valid STS signals. They are $\mathbf{c}$ in (10) and $\mathbf{d}$ in (11). All others (such as $\mathbf{c}'$ and $\mathbf{d}'$) are invalid STS signals. Therefore, there is no ambiguity in detecting $\mathbf{c}$ and $\mathbf{d}$ at the receiver.

It is easy to verify that at least $\left\lceil \dfrac{11}{2} \right\rceil = 6$ out of the 11 tones in any of the 2048 combinations belong to either $\mathbf{c}$ or $\mathbf{d}$. According to the maximum distance separable property of the STS signal, two tones

($K=2$) are sufficient to determine an STS signal. Since $\left\lceil \frac{11}{2} \right\rceil = 6 > 2$, none of the 2048 combinations except the two corresponding to STS signals **c** and **d** forms a valid STS signal. The $2048 - 2 = 2046$ combinations are nothing but different combinations of **c** and **d**. They cannot be combinations from any other (valid) STS signals. We can therefore come to the following remark as a result from Proposition 2:

*Remark 1: Assume $G$ $(G \leq D^K)$ distinctive STS signals coded on GF(D) with code rate $(N,K)$, $N \geq K$, are simultaneously received without tone detection error. Among the $G^N$ possible STS tone combination sequences, only G are valid STS signals. The rest $G^N - G$ tone sequences are simply different combinations from the G valid STS signals. They can neither form any valid STS signals nor be combinations of any other (valid) STS signals.*

However, the case of particular interest is $K=1$. When $K=1$, (9) holds for *any* value of $G$ regardless the value of *N*. We therefore have the following important remark:

*Remark 2: Assume $G$ ($G \leq D$) distinctive STS signals coded on GF(D) with code rate $(N,1)$, $\forall N \geq 1$ are simultaneously received. Under perfect tone detection, all G STS signals can be decoded to the original information symbols without ambiguity.*

Users transmitting distinctive STS signals do not cause ambiguity at a receiver. For example, (8, 1) coded STS signals in theory can support up to *D* number of simultaneous STS signal transmissions. That is, up to *D* simultaneously received STS signals can be faithfully recovered at the receiver. This is true whether the value of *N* equals 8 or not[1].

However, as alluded to above, this conclusion is only true under the assumption of ideal tone detection. In practical scenarios when STS tone detection is not error-free due to channel fading and noise, the value of *N* does affect the STS signal's capability of correcting tone detection errors. The relationship between *N* and the capability of STS tone false detection correction and missed detection recovery is given by (8) in Section A. Therefore, as long as the number of errors of STS tone detection are within the capability of the coded STS signals governed by the value of $N$, i.e., $2\varepsilon + v \leq N - 1$, multi-user ambiguity can be eliminated.

---
[1] Note that when *N=K=*1, the STS signal reduces to the frequency shift keying signal.

*Remark 3:* Assume $G$ $(G \leq D)$ *distinctive STS signals coded on GF(D) with code rate* $(N,1), \forall N \geq 1$ *are simultaneously received. All G STS signals can be decoded to their original information symbols without ambiguity as long as the numbers of tone false detection and/or missed detection satisfy*

$$2\varepsilon + v \leq N - 1 \tag{14}$$

where $\varepsilon$ and $v$ are the number of false detections and missed detections of STS tones, respectively.

Another important and challenging issue that cannot be overlooked in over-the-air signaling design for femtocell networks is sensitivity to time and frequency synchronization errors. Time and frequency among femtocells may not be as perfectly synchronized as in macrocells. Therefore, users who are synchronized to their serving base stations may not be synchronized to their neighboring base stations. Since the STS signal is meant to be received by multiple neighboring interfering base stations, the STS signal has to be designed with time and frequency offset tolerance. The time offset among femtocells is easily absorbed by the cyclic prefix of the OFDM symbol [23] and is thereby less of a concern. The design of STS with frequency offset immunity is more challenging since the information carried by an STS signal lies in the subcarrier index of the STS tones. If not appropriately designed, the STS tones may shift to neighboring subcarriers as a result of frequency offset. Again, we rely on a special coding scheme to provide frequency offset immunity.

Assume that the transmitted STS signal is

$$\mathbf{c} = \begin{bmatrix} c_1 & c_2 & \cdots & c_N \end{bmatrix} \tag{15}$$

From (2), the inverse transform of (15) is given by

$$\mathbf{Z}^{-1}\mathbf{c}^T = \begin{bmatrix} 0 & \mathbf{u}^T & 0 & \cdots & 0 \end{bmatrix}^T \tag{16}$$

with the first element equal to zero. Therefore, the inverse of a valid code word always has a zero-valued first element by design.

However, all the tones of the STS signal, transmitted by a user $k$ and received by a neighbor femtocell base station $j$ with frequency offset $\delta_{jk}$ relative to user $k$, are shifted by the same amount $\delta_{jk}$, i.e.,

$$\mathbf{c}' = \begin{bmatrix} c_1 + \delta_{jk} & c_2 + \delta_{jk} & \cdots & c_N + \delta_{jk} \end{bmatrix} \tag{17}$$

The first element of the inverse transform of (17) produces

$$\frac{1}{N}\sum_{n=1}^{N}\left(c_n + \delta_{jk}\right) = \frac{1}{N}\sum_{n=1}^{N}c_n + \delta_{jk} = \delta_{jk} \tag{18}$$

which is non-zero (for $\delta_{jk} \neq 0$). We therefore have the following remarks:

*Remark 4: An STS signal received with a frequency offset does not correspond to a valid STS signal.*

This property ensures that a receiver with a frequency offset to its transmitter will not erroneously map an STS signal to a valid (but wrong) message.

*Remark 5: The value of the first element of the inverse transform of a received STS signal equals the frequency offset between the sender and the receiver of the STS signal.*

This property enables the receiver to detect the frequency offset, if any, between the sender and the receiver during the decoding process. The frequency-offset STS signal can then be correctly recovered. This property is hence particularly beneficial to femtocell networks where base stations deployed in homes may not be perfectly frequency synchronized to each other.

*C. Design Example and Performance of STS*

In this sub-section, we provide a concrete STS design example for use in a femtocell network. We will adopt the LTE framework [32] and focus on the downlink in the following discussion. As earlier stated, STS is used for IM in femtocell networks. A user who is experiencing severe interference from one or more base stations, due to, e.g., entering the coverage of femtocell base stations with restricted access, broadcasts the interference coordination request message (ICRM) via STS to all neighboring base stations. The base stations that are able to decode the ICRM (therefore, major interferers) can simply clear the resource (no transmission on this resource) based on certain criterion (i.e. ON/OFF power control based on user's traffic data priority) or can coordinate spatial beams according to the received ICRM conveyed through STS. ON/OFF power control and spatial beam coordination will be discussed in the subsequent section.

The information included in the ICRM, in general, can be the assigned radio resource identification by the serving base station, and traffic priority indicator, etc. Specifically, the assigned radio resource identification (ID) represents the unique resource $r$, $1 \leq r \leq R$. The number of bits of this ID depends on the system bandwidth and the granularity of the resource. For the system with $D$=512 subcarriers, a 2-bit resource ID are typically sufficient to represent four ($R=4$) unique sub-bands.

Data traffic priority represents the current traffic priority of the user [33] and should be taken into account by IM to guarantee users' QoS. The traffic priority is a metric that is a function of the type of traffic flow (e.g., best effort, delay sensitive QoS flows), packet delay, queue length, current average rate and the target rate, etc. We allocate 3 bits from the ICRM to represent eight levels of data traffic priorities.

In addition, the base station ID can be programmed into the ICRM to identify the STS signals sent from users served by different cells. However, base station ID (typically 9 bits) is too long to fit into an ICRM payload. A time-varying hash function can then be used to convert the 9-bit ID into a 4-bit time-varying identification number. (Note that the ICRMs sent by users within a cell will not collide since their resource IDs are guaranteed to be distinctive by the base station scheduler). Therefore, the interference coordination request message $m$ in an STS consists of 9 information bits, i.e., 2 bits of assigned radio resource ID, 3 bits of traffic priority, and 4 bits of hashed serving base station ID. With $D=512$ subcarriers, the 9 bits information $m$ can be represented with $K=1$ information symbols according to (1) and can be further encoded into a codeword with block length $N=11$ (one LTE sub-frame excluding the first three OFDM symbols for control channels). In this setup, ideally, we can see that the inequality (9) holds for any value of *G* up to the total number of the code words. This means, as long as the numbers of false tone detections and missed tone detections satisfy $2\varepsilon + v \leq 10$, up to 512 coded STS signals can be sent simultaneously without interfering with each other or causing ambiguity at the receiver.

Fig. 5 shows the STS signal erasure and error performance in a multi-user scenario, in which the number of simultaneously transmitted STS signals from different users is *G*=30, total information bits of signaling is 9 bits, the number of subcarriers used for STS in an OFDM symbol is 512, code rate of STS is (11,1) and the number of receive antennas is 1, 2 or 4. Although in a femtocell network the number of major interferers is typical less than 10, we used 30 different STS signals to stress test the STS scheme. Performance in AWGN channel is plotted as a reference. SIR is defined as the ratio of received energy per sample (i.e., OFDM symbol time-domain sample) to interference (other users' uplink data) power plus noise (thermal noise) variance in time domain. An STS signal erasure is defined as the event in which the base station fails to decode the STS signal sent from a user, while an error is an event in which the base station decodes the STS signal to a wrong but valid STS signal. An erasure causes the base station to fail to respond to the interference coordination request whereas an error causes a base station's incorrect response to the request and may result in waste of resource. In Fig. 5, the error rate is controlled below 1%. It is observed that STS can operate at very low SIR under multi-signal (e.g., *G*=30) simultaneous transmissions. Multiple receive antennas help minimize the fading effect due to the spatial diversity resulting in performance closer to AWGN channel for the four antenna case. In the low SIR region, the up-fades from the frequency selective fading in PedB channel [34] create more opportunities than AWGN channel for the STS tones to be detected, causing fewer erasures.

As earlier stated, since the transmission of coded STS is overlaid on top of other users' uplink data transmissions, the interference caused by the transmission of STS to other users is unavoidable. However, since the interference caused by STS is isolated and STS tones are usually much stronger than regular data tones, the STS tones can be easily detected and zeroed out. In addition, isolated erasure/error can be effectively removed by the decoder. Minor effect on decoding performance is expected.

Fig. 6 shows the effect of multiple STS signals transmissions on uplink data decoding performance (Packet Error Rate). We can see that the effect on data decoding performance is less than 0.3 dB for 30 simultaneous STS signals.

## III. DISTRIBUTED DYNAMIC INTER-CELL INTERFERENCE MANAGEMENT

From the previous section, we can see that the good properties (i.e. no near-far effect, no inter-signal interference) and superior performance of the STS signal provide an effective means for applying different kinds of IM schemes. In this section, we propose two STS-based IM schemes, i.e., an ON/OFF power control scheme and a user-traffic-prioritized SLNR-based transmitter beam coordination scheme for IM in femtocell networks.

### A. System Model

For the sake of notational convenience in the following discussion, users within the same femtocell will not use the same resource, but users in different femtocells may use the same resource. Thus on resource $r$ $(1 \leq r \leq R)$, the femtocell base station and the user have a one to one relationship, i.e., on resource $r$, user $k$ is served by base station $k$, where $R$ is the total number of resources. Restricted access is assumed. Base station $i$ is equipped with $N_i$ transmit antennas and each user has one receive antenna. Let $\mathbf{h}_{ij}^r \in \mathbb{C}^{1 \times N_j}$ denote the channel between base station $j$ and user $i$ on resource $r$, whose element, $\mathbf{h}_{ij}^r(m)$, $1 \leq m \leq N_j$, represents the channel gain between the $m$th antenna of base station $j$ and the antenna of user $i$ on resource $r$, and is modeled as independent and identically distributed complex Gaussian variable with zero mean and unit variance.

We adopt the LTE TDD framework and focus on the downlink IM scenario in the following discussion.

The proposed IM approaches are illustrated in Fig. 7. There are four users competing for the same downlink resource (Fig. 7 (a)). Each user broadcasts ICRM via STS (Fig. 7 (b)) requesting for coordination from neighboring interfering base stations. The base stations that have received ICRM(s) can simply clear the resource (no transmission on the requested resource) according to certain criterion, e.g., the user traffic priority, as in the ON/OFF power control scheme; or coordinate transmit beam, as shown in Fig. 7 (c), as in the SLNR-based transmitter beam coordination scheme. Note that user 1 can not receive service from base station 4 due the restricted access although user 1 is closer to base station 4. In this case, it is clear that the coordination signals from other users would have been blocked by user 4 at base station 4 due to the near-far effect if other signaling schemes (e.g., CDMA) were used.

## B. STS-based ON/OFF Power Control

ON/OFF power control is the simplest way for managing interference using STS. Suppose that a base station $k$, which receives $U_k^r \left( U_k^r \geq 0 \right)$ ICRMs from $U_k^r$ users served by neighboring base stations and its own user requesting for resource $r$ $(1 \leq r \leq R)$, it compares the traffic priority $p_k^r$ of its served user with the priorities $\{ p_i^r \, | \, i \neq k, 1 \leq i \leq U_k^r \}$ of the competitors included in the $U_k^r$ received ICRMs. If $p_k^r < \max_{i \neq k, 1 \leq i \leq U_k^r} \{ p_i^r \}$, i.e., the served user traffic priority of the current base station is lower than its competitors, the base station simply gives up the use of resource $r$ for the next given $F$ (e.g., 1) subframe(s). If, on the other hand, $p_k^r > \max_{i \neq k, 1 \leq i \leq U_k^r} \{ p_i^r \}$, the base station ignores the ICRM requests and goes ahead using resource $r$ for the next $F$ subframes for its own user traffic. For the tied priority case, the base station simply "tosses a coin" to break the tie. This mechanism can protect bursty application users in poor channel condition from being jammed by neighboring base stations serving users with lower traffic priorities. From Fig. 8, we can see that the performance of the users under poor channel condition (experiencing severe interference from neighboring cells) has been significantly improved by ON/OFF power control compared to the case where no ICRM is applied.

## C. STS-based Transmitter Beam Coordination

In the previous subsection, the proposed ON/OFF power control scheme improves the performance of the users in poor channel conditions and thus improves the fairness of the system. However, this improvement comes at the expense of other users' bandwidth. In this subsection, we consider the potential application of the SLNR scheme [35] for transmitter beam coordination among competing femtocell base stations. The transmitter beam coordination proposed in this subsection is a more efficient way for IM by taking full advantage of the STS/ICRM from competing users. In particular, the SLNR-based precoding can be readily used for transmit beam coordination since the channel information needed for SLNR-based precoding can be easily obtained from the STS signals.

Assume that the data symbol to be transmitted on downlink resource $r$ from base station $k$ to user $k$ is $s_k$, and satisfies the power constraint $\mathbb{E}(|s_k|^2) = 1$. Before being transmitted, the data symbol $s_k$ is pre-multiplied by a precoding/beamforming vector $\mathbf{v}_k^r \in \mathbb{C}^{N_k \times 1}$ ($\|\mathbf{v}_k^r\|^2 = 1$). The total received signal at user $k$ can be written as

$$y_k^r = \sum_{i=1}^{U_k^r} \mathbf{h}_{ki}^r \mathbf{v}_i^r s_i + n_k^r$$
$$= \mathbf{h}_{kk}^r \mathbf{v}_k^r s_k + \sum_{i=1, i \neq k}^{U_k^r} \mathbf{h}_{ki}^r \mathbf{v}_i^r s_i + n_k^r \tag{19}$$

where $U_k^r$ is the number of users (including user $k$) whose ICRMs requesting for resource $r$ are successfully received by base station $k$, and $n_k^r$ is the receiver thermal noise plus other downlink data transmissions on resource $r$ that are not included in $U_k^r$ and is modeled as additive complex white Gaussian noise with zero mean and variance $\sigma_k^2$.

The original SLNR in [35] is defined as the ratio of the power of the desired transmit signal (to the transmitter's target receiver) to the power of the leakage signal (interference to other users, i.e., the victims) plus noise, the SLNR for user $k$ competing for resource $r$ can be written as:

$$SLNR_k^r(\mathbf{v}_k^r) = \frac{\left\|\mathbf{h}_{kk}^r \mathbf{v}_k^r\right\|^2}{\sigma_k^2 + \sum_{i=1, i \neq k}^{U_k^r} \left\|\mathbf{h}_{ik}^r \mathbf{v}_k^r\right\|^2}. \tag{20}$$

Note that $\mathbf{h}_{ik}^r, 1 \leq i, k \leq U_k^r$ is the downlink channel from base station $k$ to user $i$ on resource $r$. The optimal beam $\mathbf{v}_k^r$ to user $k$ at base station $k$ is the maximizer of (20). SLNR clearly makes better sense than the conventional beamforming that only optimizes the SNR (signal to noise ratio) at user $k$, i.e., $SNR_k^r(\mathbf{v}_k^r) = \frac{\left\|\mathbf{h}_{kk}^r \mathbf{v}_k^r\right\|^2}{\sigma_k^2}$, which does not take interference to other users into consideration (no coordination). However, the challenge of applying SLNR to *inter*-cell interference management is that it requires the *interference/leakage channel* knowledge, $\mathbf{h}_{ik}^r, 1 \leq i \leq U_k^r, i \neq k$, which is the channel from the $k$th femtocell to user $i$ served by a *different* cell $i$ $(i \neq k)$. This knowledge is thus typically not available.

Due to the reciprocity property of the channel, $\mathbf{h}_{ik}^r = (\mathbf{g}_{ki}^r)^{\mathrm{T}}$, $1 \leq i, k \leq U_k$, where $\mathbf{g}_{ki}^r$ is the uplink channel on resource $r$ from user $i$ to base station $k$, whose $n$th element $\mathbf{g}_{ki}^r(n)$, $1 \leq n \leq N_k$ is the uplink channel gain on resource $r$ from the antenna of user $i$ to antenna $n$ of base station $k$. With the STS scheme, the uplink channel from user $i$ to base station $k$, $\mathbf{g}_{ki}^r$, can be estimated at base station $k$ using the STS tones transmitted by user $i$ as the reference signals/pilots [36][37]. The optimal beamformer targeted to user $k$ can then be calculated by maximizing (20).

However, the original SLNR objective function (20) does not reflect the user traffic priorities. In fact user traffic is treated equally. QoS is thus hard to be finely controlled. In order to take individual user traffic priority into consideration during the beam coordination process, we propose a *user-traffic-prioritized* SLNR algorithm based on STS and the ICRM carried by STS for transmitter beam control. Specifically, we incorporate the data traffic priority available from the ICRM into the SLNR objective function, that is,

$$SLNR_k^r(\mathbf{v}_k^r) = \frac{\left\|\mathbf{h}_{kk}^r \mathbf{v}_k^r\right\|^2}{\sigma_k^2 + \sum_{i=1, i \neq k}^{U_k^r} \left\|p_i^r \mathbf{h}_{ik}^r \mathbf{v}_k^r\right\|^2}, . \tag{21}$$

where $p_i^r$ corresponds to the traffic priority for user $i$ (served by base station $i$) who is competing for resource $r$. It is clear that (21) provides a transmit beam coordination metric with an additional mechanism for fine tuning the traffic QoS. The traffic-prioritized SLNR precoder can thus be obtained by maximizing (21). Using the Rayleigh-Ritz quotient and dropping the resource index $r$ (for notational simplicity) yields:

$$\mathbf{v}_k^* \propto \max \text{ eigenvector}\left(\left(\sigma_k^2 \mathbf{I} + \left(\mathbf{P}_k \mathbf{H}_k\right)^H \mathbf{P}_k \mathbf{H}_k\right)^{-1} \mathbf{h}_{kk}^H \mathbf{h}_{kk}\right), \tag{22}$$

where

$$\mathbf{H}_k = \begin{bmatrix} \mathbf{h}_{1k} & \mathbf{h}_{2k} & \cdots & \mathbf{h}_{(k-1)k} & \mathbf{h}_{(k+1)k} & \cdots & \mathbf{h}_{U_k k} \end{bmatrix}^T \tag{23}$$

is the *leakage channel* corresponding to user $k$ (served by base station $k$), which can be estimated from the corresponding STS signals transmitted from the users of neighboring cells $\{i, i \neq k\}$ (i.e., victims of base station/user $k$) received at base station $k$, and

$$\mathbf{P}_k = diag\left(p_1 \ p_2 \ \cdots \ p_{k-1} \ p_{k+1} \cdots \ p_{U_k}\right) \tag{24}$$

is the *victim traffic priority matrix* for user $k$.

Fig. 9 shows the performance of the proposed STS based beam coordination schemes versus that without ICRM as well as the conventional ideal beamforming scheme (no coordination). Like the ON/OFF power control scheme, the proposed beam coordination schemes significantly improve the performance of the users with poor channel condition (experiencing severe interference from neighboring cells), however, without costing the good (channel condition) user's performance. In addition, the traffic-prioritized SLNR beam coordination allows for further fine-tuning of the poor channel condition users' performance in accordance to the users' *current* traffic priorities.

## IV. CONCLUSIONS

In this paper, the problem of inter-cell interference management (IM) for interference-dominated femtocell networks is studied. Two fast IM techniques using over-the-air signaling are proposed. The enabler of these techniques is the use of a special signal, i.e., the single-tone signal. The proposed single-tone signaling (STS) scheme has many most desirable properties. First, the STS scheme does not require dedicated resource, i.e., STS overlays on top of the other users' uplink data transmissions and thereby does not incur system overhead. The STS *does* however cause interference to other users' uplink data transmissions. However, since, unlike the conventional CDMA signals whose energy is spread over a contiguous bandwidth, STS energy is concentrated only on one subcarrier per OFDM symbol, the effect on the other users' uplink data decoding is minor. In fact since STS tones are easy to be detected, the effect can be further minimized by removing (zeroing-out) the subcarrier where a strong STS tone is detected. Second, the STS tones from different STS signals are indistinctive, therefore, have no near-far effect. That is, users transmitting STS in the neighboring cells will not be blocked by the local cell users who are broadcasting their own STS signals, thereby significantly increasing the hearability of the STS signals. The indistinctive property of the STS tones therefore prevents interference among users, however, it results in ambiguity of STS signals among users at the receiver. This ambiguity is resolved by encoding (non-binary) the STS tone positions using Galois field transform. Third, the STS signal is a single tone waveform with low PAPR, which is especially beneficial for coverage extension and mobile device power amplifiers. Finally, STS is robust to synchronization errors that are crucial to *inter*-cell communications. Based on the STS scheme, the ON/OFF power control and the SLNR-based transmitter beam coordination schemes are proposed. The ON/OFF power control scheme utilizes interferees' traffic priority information conveyed by STS for IM. Whereas the SLNR-based scheme uses the *interference* channel information provided by the STS for beam coordination among different cells. In addition, the user-traffic-prioritized SLNR-based beam coordination scheme provides QoS as well as fairness without the necessity for a centralized scheduler, thus it's a distributed scheme. The proposed IM schemes provide an effective means for fast rescuing users who are experiencing severe interference from neighboring femtocells due to, e.g., restricted access.

REFERENCES

[1] D. N. Knisely, T. Yoshizawa, and F. Favichia, "Standardization of femtocells in 3GPP," *IEEE Commun. Mag.*, vol. 47, no. 9, pp. 68-75, Sep. 2009.

[2] R. Kim, J. Kwak, and K. Etemad, "WiMAX femtocell: requirements, challenges, and solutions," *IEEE Commun. Mag.*, vol. 47, no. 9, Sep. 2009.

[3] V. Chandrasekhar, J. G. Andrews, T. Muharemovic, Z. Shen, and A. Gatherer, "Power control in two-tier femtocell networks," *IEEE Trans. Wireless Commun.*, vol. 8, no. 8, pp. 4316-4328, Aug. 2009.

[4] V. Chandrasekhar, J. G. Andrews, "Spectrum allocation in tiered cellular networks," *IEEE Trans. Commun.*, vol.57, no.10, pp.3059-3068, Oct. 2009.

[5] S. Park, W. Seo, S. Choi, D. Hong, "A Beamforming Codebook Restriction for Cross-Tier Interference Coordination in Two-Tier Femtocell Networks," *IEEE Trans. Veh. Technol.*, vol.60, no.4, pp.1651-1663, May 2011.

[6] H.-S. Jo, C. Mun, J. Moon, and J.-G. Yook, "Interference mitigation using uplink power control for two-tier femtocell networks," *IEEE Trans. Wireless Commun.*, vol. 8, no. 10, pp. 4906-4910, Oct. 2009.

[7] J. W. Huang, V. Krishnamurthy, "Cognitive Base Stations in LTE/3GPP Femtocells: A Correlated Equilibrium Game-Theoretic Approach," *IEEE Trans. Commun.*, vol.59, no.12, pp.3485-3493, Dec. 2011.

[8] V. Chandrasekhar and J. G. Andrews, "Uplink capacity and interference avoidance for two-tier femtocell networks," *IEEE Trans. Wireless Commun.*, vol. 8, no. 7, pp. 3498-3509, Jul. 2009.

[9] Y.-S. Liang, W.-H. Chung, G.-K. Ni, I.-Y. Chen, H. Zhang, and S.-Y. Kuo, "Resource Allocation with Interference

Avoidance in OFDMA Femtocell Networks," *IEEE Trans. Veh. Technol.*, vol.61, no.5, pp.2243-2255, Jun. 2012.

[10] D. Lopez-Perez, A. Valcarce, G. de la Roche, and J. Zhang, "OFDMA femtocells: A roadmap on interference avoidance," *IEEE Commun. Mag.*, vol. 47, no. 9, pp. 41-48, Sep. 2009.

[11] P. S. Rha, "Frequency reuse scheme with reduced co-channel interference for fixed cellular systems," *Electron. Lett.*, vol. 34, no. 3, pp. 237-238, Feb. 1998.

[12] 3GPP TS 36.420 V10.1.0, 3rd Generation Partnership Project; Technical Specification Group Radio Access Network; Evolved Universal Terrestrial Radio Access Network (E-UTRAN); X2 general aspects and principles (Release 10), Jun. 2011.

[13] V. Chandrasekhar, J. G. Andrews, and A. Gatherer, "Femtocell networks: a survey," *IEEE Commun. Mag.*, vol. 46, no. 9, pp. 59-67, Sep. 2008.

[14] P. Zhou, H. Hu, H. Wang, and H.-H. Chen, "An efficient random access scheme for OFDMA systems with implicit message transmission," *IEEE Trans. Wireless Commun.*, vol. 7, no. 7, pp. 2790-2797, Jul. 2008.

[15] C.-S. Hwang, K. Seong, and J. Cioffi, "Throughput maximization by utilizing multi-user diversity in slow-fading random access channels," *IEEE Trans. Wireless Commun.*, vol. 7, no. 7, pp. 2526-2535, Jul. 2008.

[16] F. Khan, *LTE for 4G Mobile Broadband: Air Interface Technologies and Performance*, Cambridge University Press, 2009.

[17] V. Aggarwal and A. Sabharwal, "Performance of multiple access channels with asymmetric feedback," *IEEE J. Sel. Areas Commun.*, vol. 26, no. 8, pp. 1516-1525, Oct. 2008.

[18] J.-P. M. G. Linnartz, R. Hekmat, and R.-J. Venema, "Near-far effects in land mobile random access networks with narrow-band Rayleigh fading channels," *IEEE Trans. Veh. Technol.*, vol. 41, no. 1, pp. 77-90, Feb. 1992.

[19] H. H. Chen, *the Next Generation CDMA Technologies*, John Wiley & Sons, 2007.

[20] H. Holma, A. Toskala, *WCDMA for UMTS: HSPA Evolution and LTE*, John Wiley & Sons, 2010.

[21] S. Sesia, I. Toufik, M. Baker, *LTE: The UMTS Long Term Evolution*, Wiley, 2011.

[22] S. Ahmadi, *Mobile WiMAX: A Systems Approach to Understanding IEEE 802.16m Radio Access Technology*, 2010.

[23] R. Prasad, *OFDM for Wireless Communications Systems*, Artech House Publishers, 2004.

[24] M. Wang, J. Borran, T. Ji, T. Richardson, M. Dong, "Interference management and handoff techniques in Modern OFDMA cellular communication systems," *IEEE Veh. Technol. Mag.*, pp.64-75, December 2009.

[25] G. Beradinelli, L. Temino, S. Frattasi, M. Rahman, and P. Mogensen, "OFDMA vs. SC-FDMA: performance comparison in local area imt-a scenarios," *IEEE Wireless Commun.*, vol. 15, no. 5, pp. 64-72, 2008.

[26] T. Jiang, M. Guizani, H. H. Chen, W. Xiang, Y. Wu, "Derivation of PAPR Distribution for OFDM Wireless Systems Based on Extreme Value Theory," *IEEE Trans. Wireless Commun.* vol.7, no.4, pp.1298-1305, Apr. 2008.

[27] T. Jiang, Y. Wu, "An Overview: Peak-to-Average Power Ratio Reduction Techniques for OFDM Signals," *IEEE Trans. Broadcast.*, vol.54, no.2, pp.257-268, Jun. 2008.

[28] S. Cripps, *Advanced Techniques in RF Power Amplifier Design*, 2002.

[29] D.Chu, "Polyphase codes with good periodic correlation properties (Corresp.),"*IEEE Trans. Inform. Theory*, vol.18, no.4, pp. 531- 532, Jul. 1972.

[30] R. Blahut, *Algebraic Codes for Data Transmission*, Cambridge University Press, 2003.

[31] Samuel R. Buss and Gyorgy Turán. "Resolution proofs of generalized pigeonhole principles." *Theoretical Computer Science,* vol.62, no.3 pp.311-317, Dec. 1988.

[32] 3GPP TR 36.912 V2.0.0, 3rd Generation Partnership Project; Technical Specification Group Radio Access Network; Feasibility Study for Further Advancements for E-UTRA (LTE-Advanced) (Release 9), Aug. 2009.

[33] A. Stolyar, "On the asymptotic optimality of the gradient scheduling algorithm for multiuser throughput allocation," *Operation Research*, vol. 53, iss. 1, pp. 12-15, 2005.

[34] Guidelines for evaluation of radio transmission technologies for IMT- 2000, ITU Radiocommunication Sector Std. Recommendation M.1225, 1997.

[35] M. Sadek, A. Tarighat, and A. H. Sayed, "A Leakage-Based Precoding Scheme for Downlink Multi-User MIMO

Channels," *IEEE Trans. Wireless Commun.*, vol. 6, no. 5, pp.1711-1721, May 2007.

[36] T. Liu, M. Wang, Y. Liang, F. Shu, J. Wang, W. Sheng, and Q. Chen, "A Minimum-Complexity High-Performance Channel Estimator for MIMO-OFDM Communications," *IEEE Trans. Veh. Technol.*, vol.59, no.9, pp.4634-4639, Nov. 2010.

[37] P. Bertrand, "Channel Gain Estimation from Sounding Reference Signal in LTE," in *Proc. IEEE 73$^{rd}$ Vehicular Technology Conf. (VTC Spring), 2011*, pp.1-5.

[38] 3GPP RAN4, "Simulation Assumptions and Parameters for FDD HeNB RF Requirements," R4-092042, May 2009.

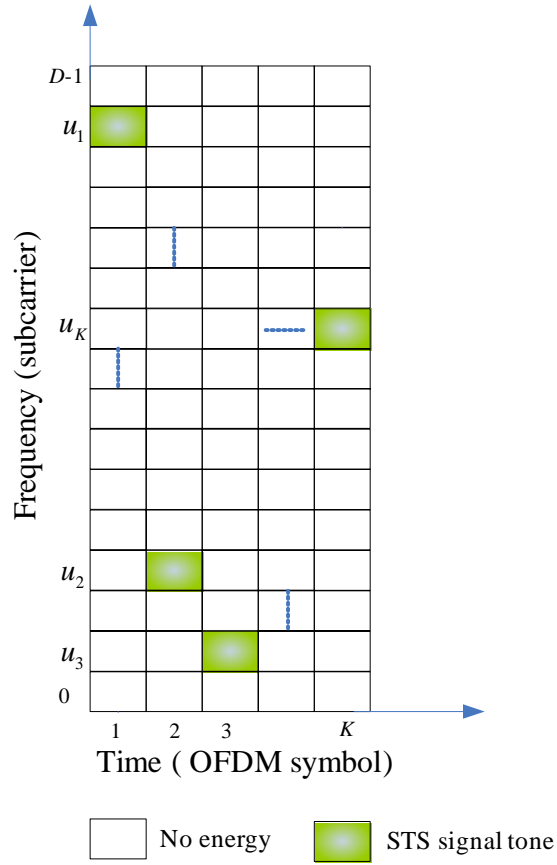

Fig. 1. Illustration of an STS signal, where information symbol of the STS signal, $u_k, 1 \leq k \leq K$, is the index of the energized subcarrier (i.e. the index of STS signal tone). Only the subcarriers used for STS signal transmission are drawn and numbered from 0 to $D-1$.

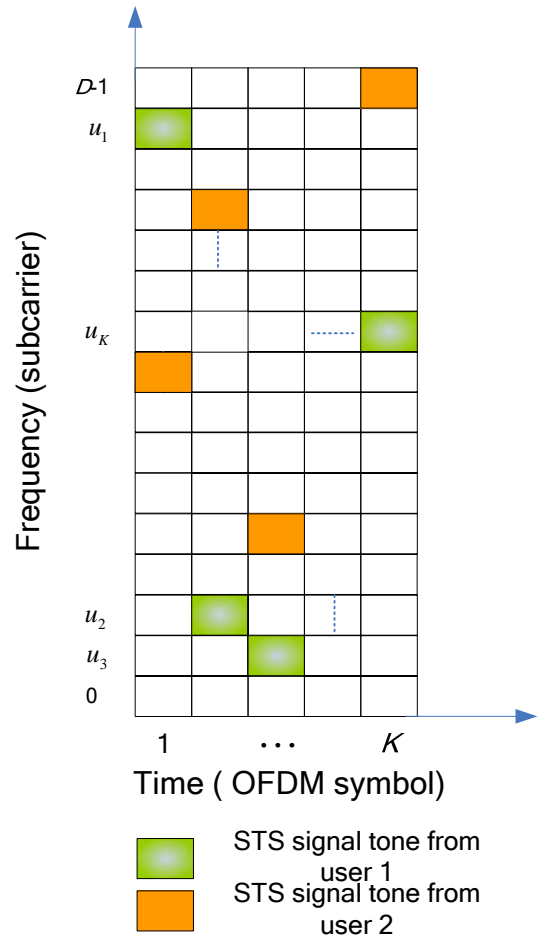

Fig. 2. Illustration of two STS signals from two users seen at a receiver (a base station). The tones themselves are not modulated. They are hence not distinctive between STS signals sent from different users although the tones are colored differently in the diagram for illustration purpose. The index for the STS signal from user 2, $v_k$ ($1 \leq k \leq K$), is not shown.

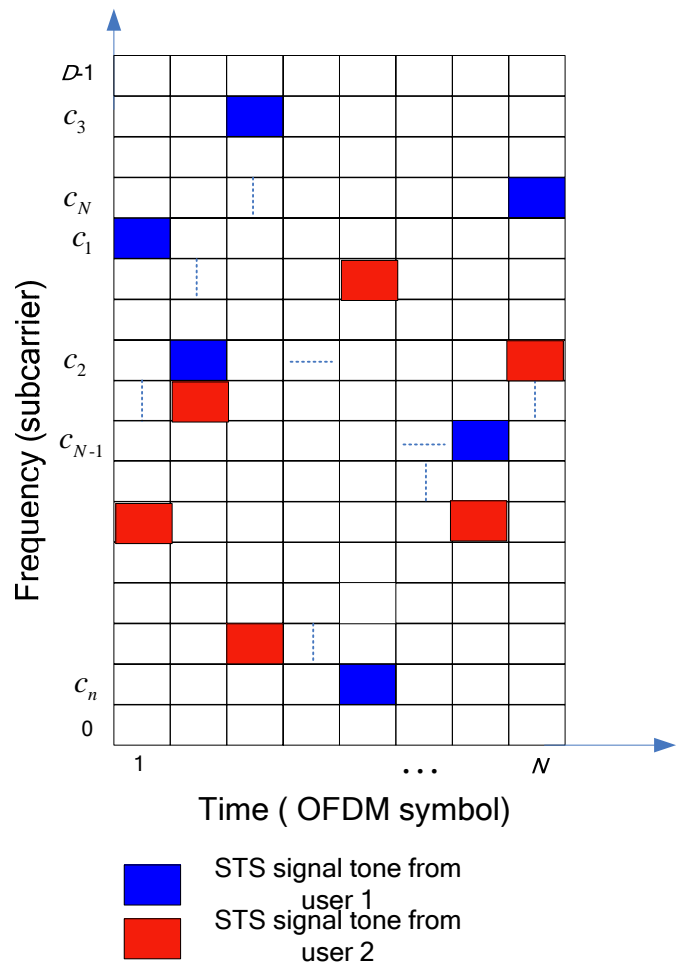

Fig. 3. Illustration of two coded STS signals from two users seen at a receiver (a base station) where the code symbol of an STS signal, $c_n, 1 \leq n \leq N$, is the index of the STS signal tone. The index for the STS signal from user 2, $d_n, 1 \leq n \leq N$, is not shown. Note the STS tones are not distinguishable between STS signals, although they are marked with different colors for illustration purpose.

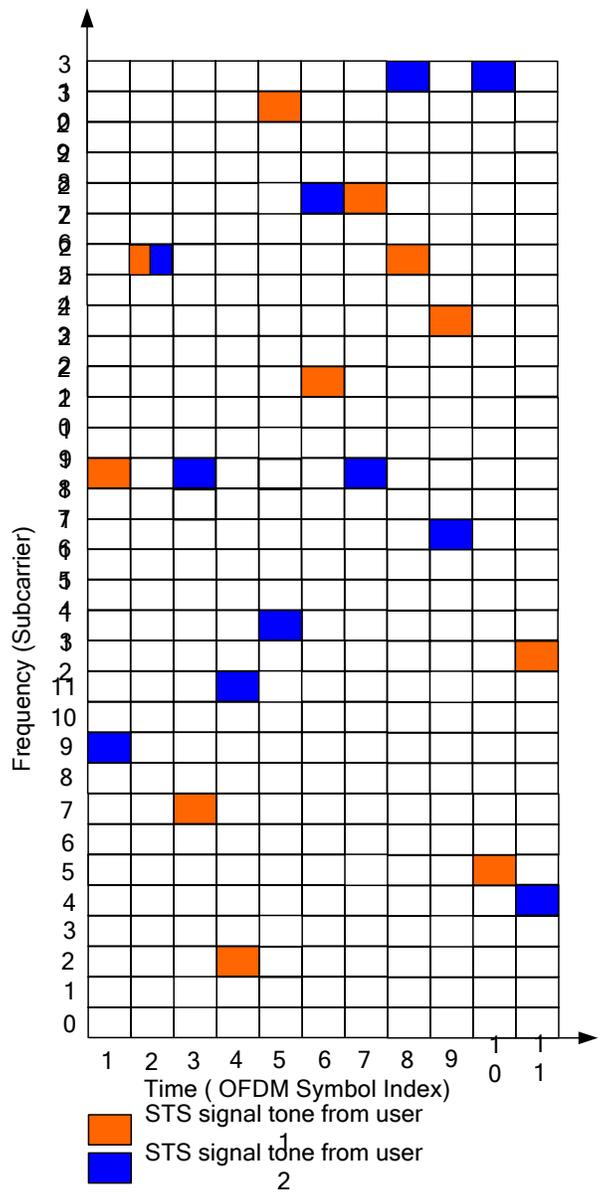

Fig. 4. An example of detecting two STS signals with code rate (11, 2) on *GF*(32).

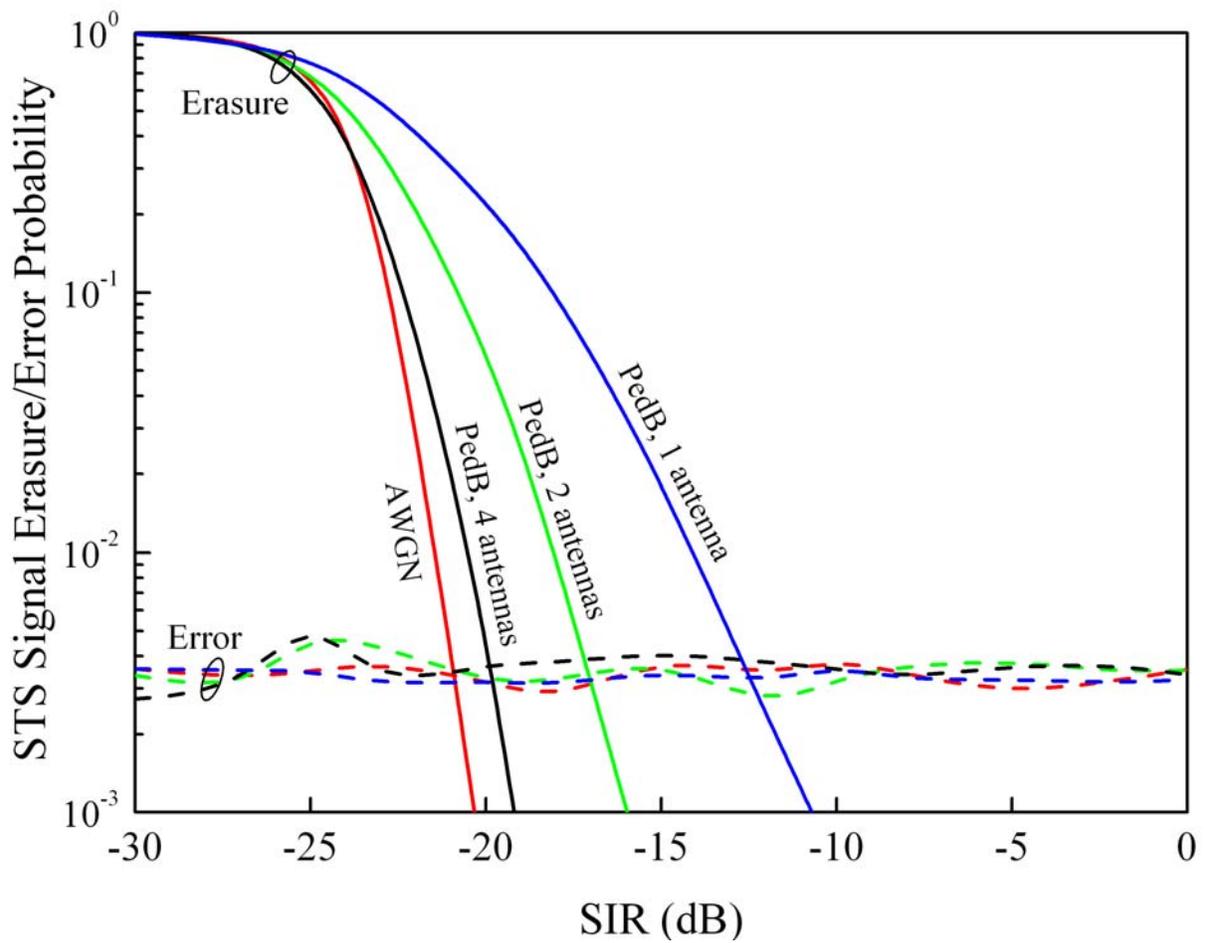

Fig. 5. The STS signal erasure and error performance in a multi-signal scenario (30 STS signals; total information bits of signaling is 9 bits; the number of subcarriers used for STS in an OFDM symbol = 512, code rate of STS $=(11,1)$, fading speed = 3 km/h at 2 GHz carrier frequency; number of receive antennas =1, 2 or 4).

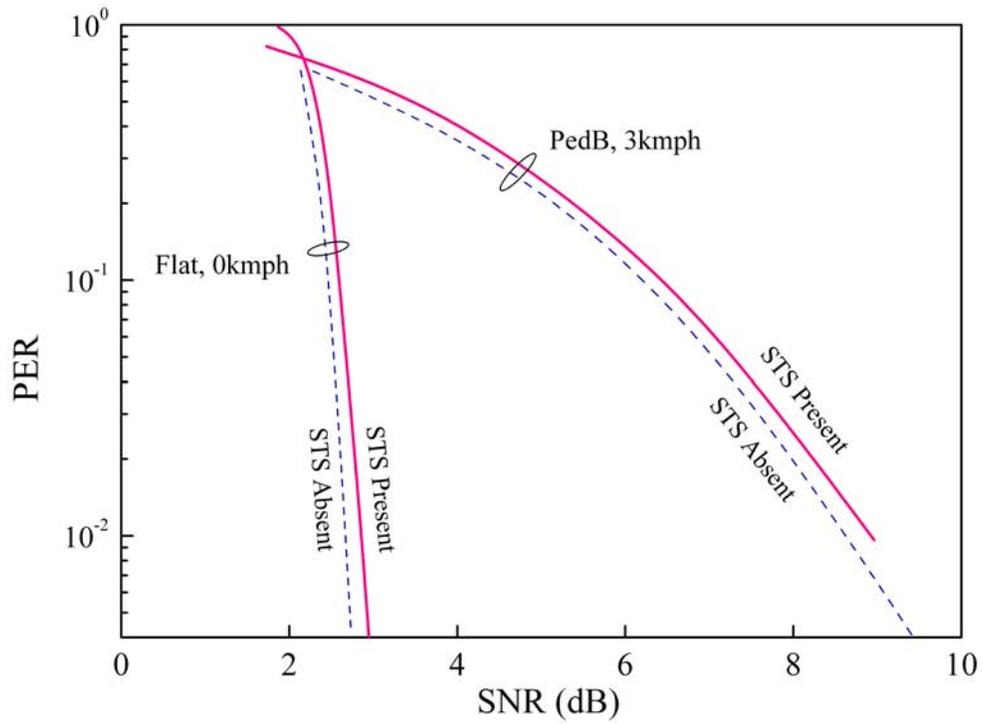

Fig. 6. Effect of multiple STS signals transmissions on uplink data decoding performance (Packet Error Rate). The number of STS signals is 30. SNR is defined as the receive tone SNR of uplink data per antenna; Uplink data modulation is 16QAM, and bandwidth is 512 subcarriers at a carrier frequency of 2 GHz.

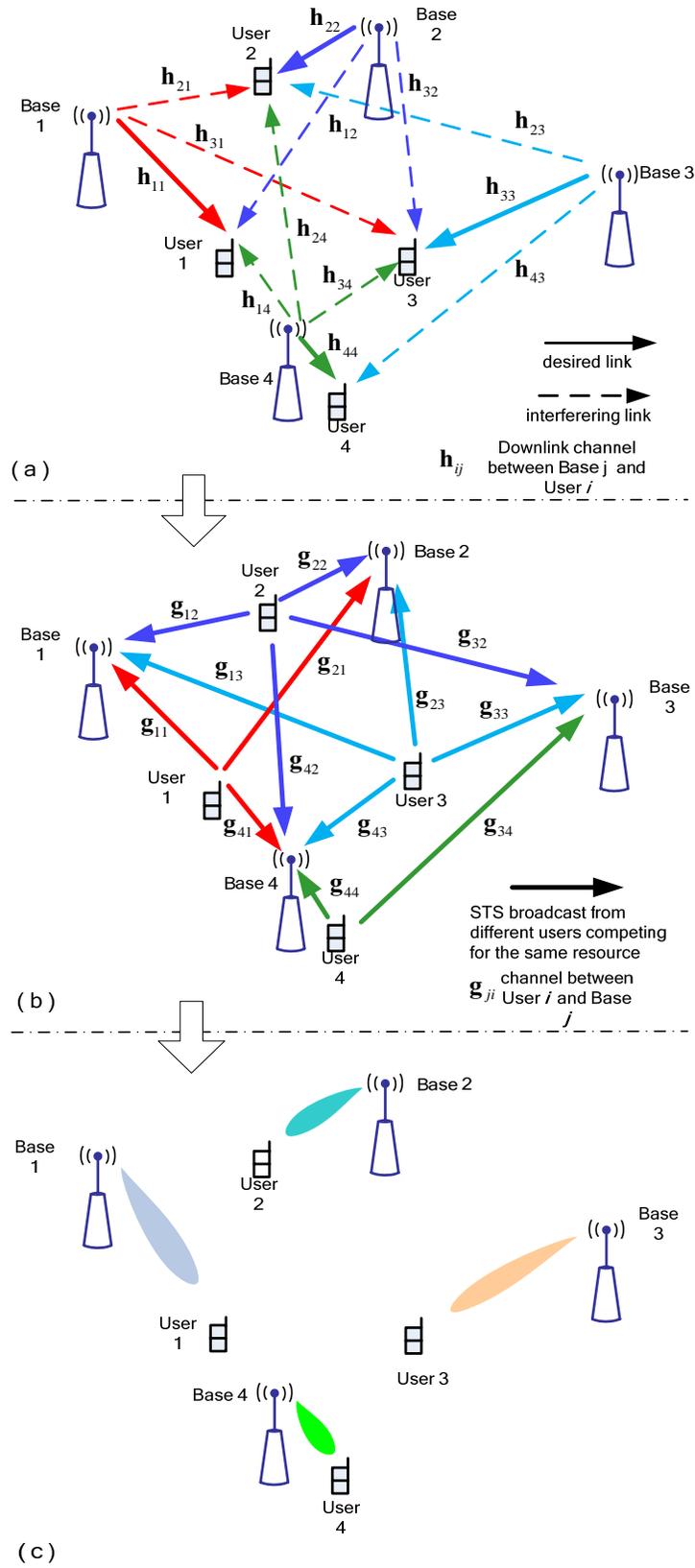

Fig. 7. Illustration of the proposed interference management (IM) schemes. Restricted access is assumed. (a) Downlink traffic interference; (b) STS transmission on uplink; (c) Transmitter beam coordination.

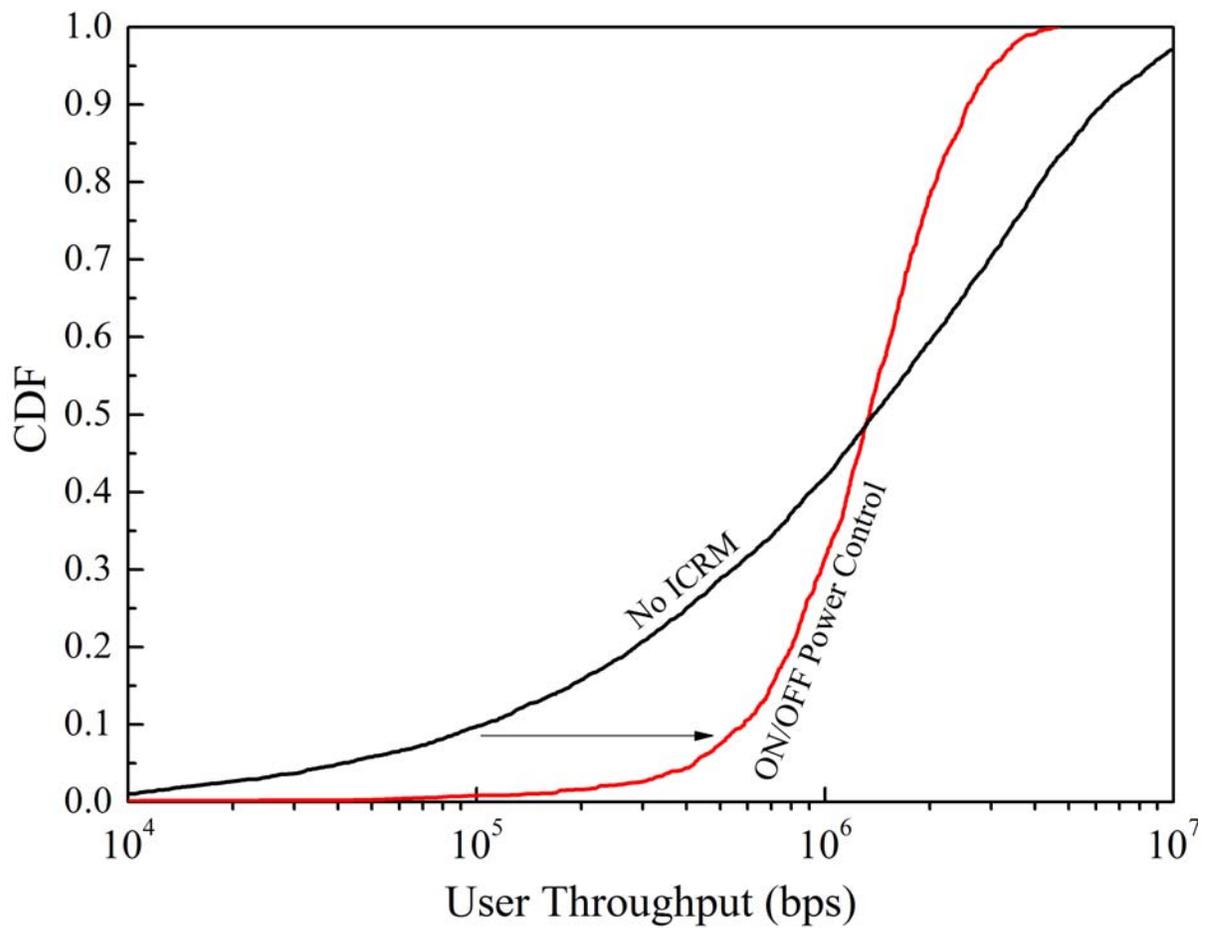

Fig. 8. Performance of STS-based ON/OFF power control, where users are repeatedly and randomly dropped in a 5×5 femtocell cluster model [38]. The disadvantageous users receive interference relief (as the arrow indicates) by broadcasting ICRM through STS.

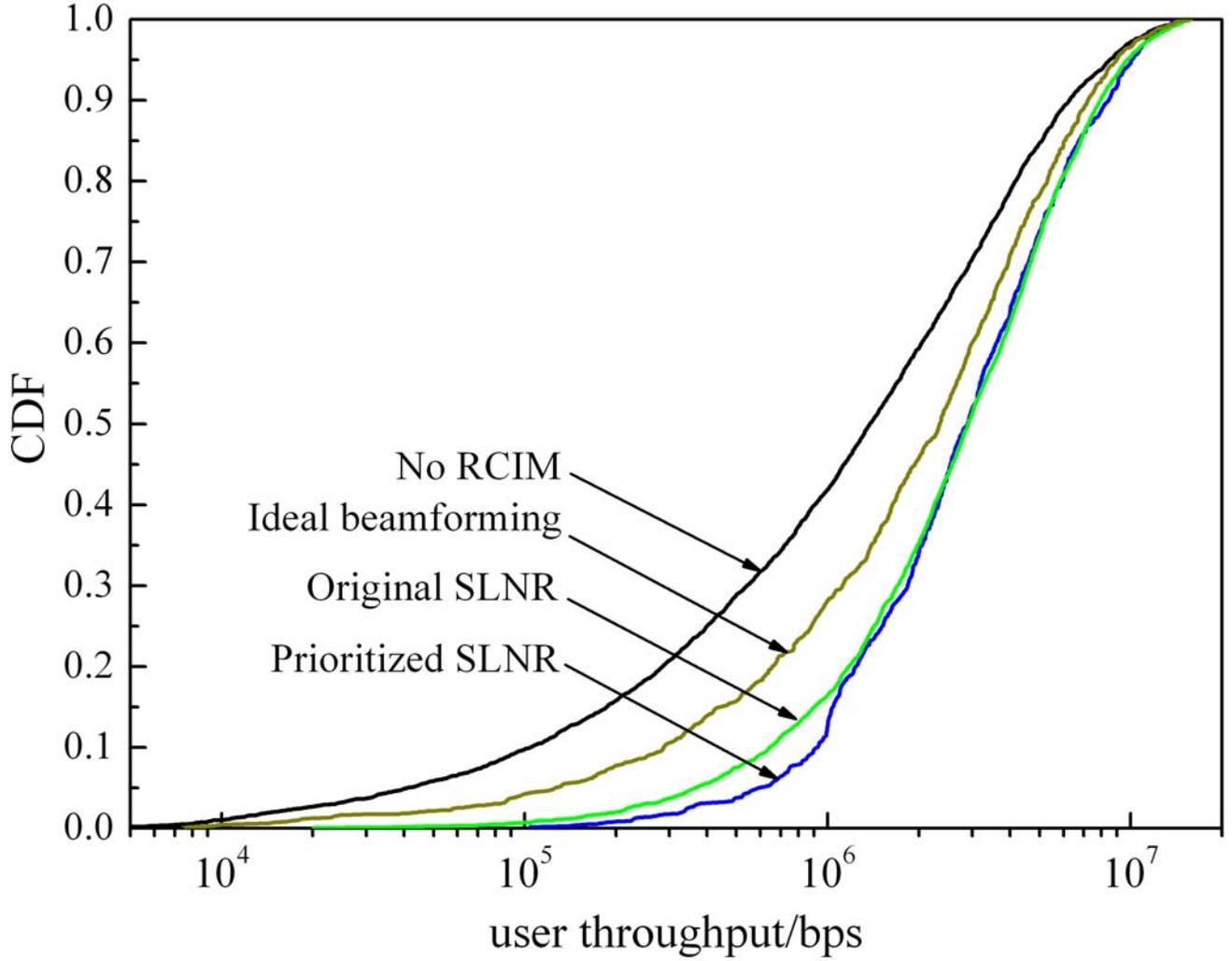

Fig. 9. Performance of the proposed STS-based beam coordination schemes. A 5×5 femtocell cluster model [38] was used. The performance without ICRM and with conventional ideal beamforming (no coordination) are plotted as performance benchmarks. The number of antennas at a base station is 2. Simulation parameters are summarized in Table. 1.

**Table. 1. Simulation parameters.**

| Deployment ratio | 0.5 | Base station receiver noise figure (dB) | 6 |
|---|---|---|---|
| Carrier frequency (GHz) | 2 | User transmit power (dBm) | 23 |
| Base station number of antennas | 2 | User receiver noise figure (dB) | 10 |
| Base station transmit power (dBm) | 23 | Number of antennas per user | 1 |
| Noise PSD (dBm/Hz) | -174 | Number of subcarriers | 512 |
| Base station noise figure (dB) | 6 | Number of bits in STS signal | 9 |
| Fading speed (kmph) | 3 | STS code rate | (11,1) |